\documentclass[aps,prb,superscriptaddress,showpacs,twocolumn]{revtex4}
\usepackage{graphicx}
\begin{document}
\bibliographystyle{apsrev}
\title{Polar catastrophe and electronic reconstructions at the LaAlO$_3$/SrTiO$_3$ interface:
evidence from optical second harmonic generation}
\author{A. Savoia}
\author{D. Paparo}
\author{P. Perna}
\author{Z. Ristic}
\author{M. Salluzzo}
\author{F. Miletto Granozio}
\author{U. Scotti di Uccio}
\affiliation{CNR-INFM Coherentia and Dipartimento di Scienze
Fisiche, Universit\`{a} di Napoli ``Federico II'', Compl.\ Univ.\ di
Monte S.Angelo, v.\ Cintia, 80126 Napoli, Italy}
\author{C. Richter}
\author{S. Thiel}
\author{J. Mannhart}
\affiliation{Center for Electronic Correlations and Magnetism,
University of Augsburg, D-86135 Augsburg, Germany}
\author{L. Marrucci}
\email{lorenzo.marrucci@na.infn.it} \affiliation{CNR-INFM Coherentia
and Dipartimento di Scienze Fisiche, Universit\`{a} di Napoli
``Federico II'', Compl.\ Univ.\ di Monte S.Angelo, v.\ Cintia, 80126
Napoli, Italy}
\date{\today}
\begin{abstract}
The so-called ``polar catastrophe'', a sudden electronic
reconstruction taking place to compensate for the interfacial ionic
polar discontinuity, is currently considered as a likely factor to
explain the surprising conductivity of the interface between the
insulators LaAlO$_3$ and SrTiO$_3$. We applied optical second
harmonic generation, a technique that \emph{a priori} can detect
both mobile and localized interfacial electrons, to investigating
the electronic polar reconstructions taking place at the interface.
As the LaAlO$_3$ film thickness is increased, we identify two abrupt
electronic rearrangements: the first takes place at a thickness of 3
unit cells, in the insulating state; the second occurs at a
thickness of 4-6 unit cells, i.e., just above the threshold for
which the samples become conducting. Two possible physical scenarios
behind these observations are proposed. The first is based on an
electronic transfer into localized electronic states at the
interface that acts as a precursor of the conductivity onset. In the
second scenario, the signal variations are attributed to the strong
ionic relaxations taking place in the LaAlO$_3$ layer.
\end{abstract}
 \maketitle


\section{Introduction}
The observation that the interface between the two band insulators
LaAlO$_3$ (LAO) and SrTiO$_3$ (STO) can be highly conducting
\cite{ohtomo04} has spurred a flourishing of research activities,
motivated both by the fundamental questions posed by this unexpected
phenomenon and by the associated technological prospects (see, e.g.,
Refs.\
\onlinecite{nakagawa06,thiel06,vonk07,siemons07,brinkman07,reyren07,willmott07,hotta07,cen08,basletic08,yoshimatsu08,caviglia08,cen09},
or Refs.\ \onlinecite{pauli08,mannhart08,huijben09} for recent
reviews). Since its initial discovery, several important features of
this puzzling phenomenon have been well established. LAO-STO
heterostructures consisting of a LAO film grown on the (001) surface
of a STO substrate are only conducting if the interface shows a
(LaO)$^+$/(TiO$_2$)$^0$ stacking, while they are insulating for
(AlO$_2$)$^-$/(SrO)$^0$ interfaces.\cite{ohtomo04} For well oxidized
systems, the former interfaces are conducting only when the
thickness of the LAO layer is at least 4 unit cells (u.c.),
otherwise they are insulating.\cite{thiel06,siemons07} Moreover, the
charge carriers are found to be localized in a interfacial layer
that is only few nanometers thick
\cite{thiel06,siemons07,basletic08} and, below $\simeq$200 mK, they
give rise to two-dimensional superconductivity.\cite{reyren07}

A leading interpretation for this interfacial conductivity is based
on the ``polar catastrophe'' mechanism (see, e.g., Ref.\
\onlinecite{nakagawa06}). The polar stacking of the charged LAO
atomic planes on the neutral STO planes gives rise to an
electrostatic potential difference across the LAO film that
increases proportionally to its thickness and hence, for
sufficiently thick LAO films, must be relaxed by an interfacial
reconstruction. The latter could be ionic, involving lattice
distortions and/or some degree of cationic
mixing,\cite{nakagawa06,vonk07,willmott07,zhong08,pentcheva09,jia09}
but it has been proposed that an \emph{electronic reconstruction}
may instead be the dominating effect, involving a transfer of
electrons from LAO to STO, likely into the STO Ti $3d$ conduction
band close to the interface, thus giving rise to the interfacial
conduction.\cite{ohtomo04,nakagawa06,thiel06,hotta07,popovic08,chen09,sing09}
Although this model seems to provide an appealing explanation for
many important features of the observed phenomena, there are issues
left unresolved, and a general consensus on the correct physical
interpretation has not been reached yet.\cite{pauli08} One example
of an unresolved issue is the difference between the electronic
carrier density measured in well oxidized samples
($2-4\times10^{13}$ cm$^{-2}$)\cite{thiel06,siemons07} and that
predicted by the polar catastrophe model ($3\times10^{14}$
cm$^{-2}$). A possible explanation for this ``missing charge''
problem is that part of the electrons injected into the interface
are localized and therefore do not contribute to the
conduction.\cite{popovic08}

To resolve this problem, and more generally to move forward in our
understanding, it is desirable to directly probe the rearrangements
of all interfacial electrons, rather than of the mobile carriers
only. \emph{Second harmonic generation} (SHG), a nonlinear optical
technique based on the detection of doubled-frequency photons in the
light reflected (or transmitted) from the interface, provides just
this capability.\cite{shen94} When the illuminated materials are
centrosymmetric, second-harmonic (SH) photons are generated with
high efficiency only in the thin interfacial regions in which the
inversion symmetry of the electronic orbitals is broken. SHG has
already been successfully applied to studying interfaces between
other perovskite oxides \cite{yamada04,ogawa08} and, concurrently to
the present work, to LAO/STO superlattices.\cite{ogawa09} Using SHG,
variations in the degree of interfacial polarity associated with
electronic reconstructions are expected to be detectable with high
sensitivity. The SHG signal can be regarded as a ``weighted
average'' of the degree of polar asymmetry felt by all electrons
present in the system, with a weight given approximately by the
electron polarizability at optical frequencies. In the present work,
we have used SHG to analyze the LAO/STO system in a set of samples
in which the thickness $d$ of the LAO layers was varied from an
undercritical thickness with insulating interfaces, through the
critical thickness, up to thick LAO layers which generate well
conducting samples.

\section{Experiment}
\begin{figure}[t]
\includegraphics[scale=0.8]{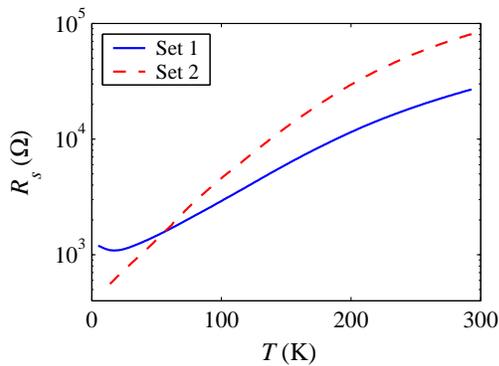}
\caption{(color online). Sheet resistance
versus temperature for two LAO/STO conducting samples having LAO
thickness $d=4$ u.c., one from set 1 (solid line, blue) and the
other from set 2 (dashed line, red). In both a metallic behavior is
evident, with a small resistance increase for the sample of set 1 at
low temperatures.} \label{fig_RT}
\end{figure}
LAO films were grown by pulsed laser deposition on STO(001)
substrates with TiO$_2$ plane termination, while controlling the LAO
thickness on a unit cell scale using high-energy electron
diffraction (RHEED) oscillations. A first set of samples (set 1,
manufactured in Naples) was grown at $\simeq$800 $^\circ$C in an
oxygen atmosphere of $1\times10^{-4}$ mbar, and then cooled at the
same pressure to room temperature. A second set of samples (set 2,
manufactured in Augsburg) was prepared in $8\times10^{-5}$ mbar of
O$_2$ at 770 $^\circ$C and cooled in 400 mbar of O$_2$. In both
sets, interfacial conduction appears only for a LAO thickness
$d\geq4$ u.c., in agreement with previous results.\cite{thiel06} Two
examples of the typical resistivity temperature-dependence of
conducting samples are shown in Fig.~\ref{fig_RT}. An additional
$d=3$ u.c.\ sample of set 2 was fabricated with a back-gate for
field-effect switching.\cite{thiel06} Hall measurements yielded
interfacial carrier densities of $\sim10^{14}$ cm$^{-2}$ in
conducting samples of set 1 and of $\sim10^{13}$ cm$^{-2}$ in those
of set 2 (at 300 K). All SHG measurements were performed at room
temperature in air and in dark (the samples were also kept in dark
for 24 hours before the measurements), after cleaning the sample
surfaces with isopropyl alcohol.

\begin{figure}[t]
\includegraphics[scale=0.6, bb= 160 230 420 635,
clip=true, angle=270]{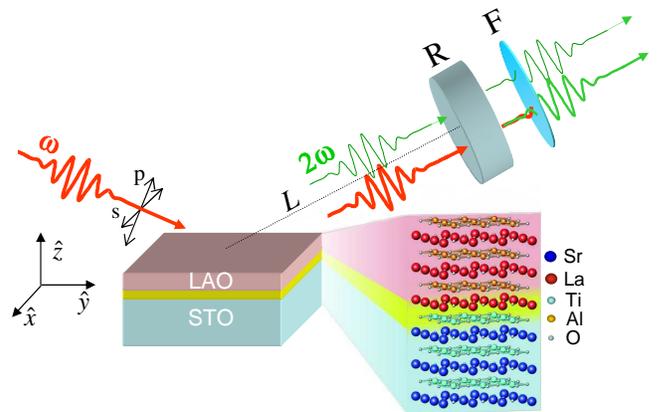} \caption{(color online). Schematic
of the homodyne SHG experiment. The sample is irradiated with laser
pulses at frequency $\omega$ (drawn in red). The SH light
($2\omega$, in green) generated in reflection by the upper surface
of the sample (including the interface) is made to interfere with
the SH generated by a reference quartz crystal (R) illuminated by
the reflected beam at the fundamental frequency (for clarity, in the
figure the two beams are shown as being spatially separated; in
reality, they are almost perfectly collinear and superimposed). The
latter is moved along the beam path (with displacement $L$), so as
to modulate the phase difference of the two SH terms by exploiting
air dispersion. A filter (F) stops the reflected light of frequency
$\omega$ before detection. The incidence angle is 64$^{\circ}$. The
input/output polarizations $s$ and $p$ used in our experiments are
also shown, with $s$ denoting an optical electric field parallel to
the sample surface $xy$, and $p$ a field lying in the incidence
plane $yz$.} \label{fig_experiment}
\end{figure}

\begin{figure}[t]
\includegraphics[scale=0.75]{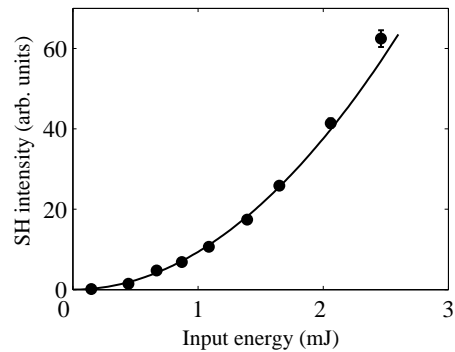}
\caption{Example of the SHG signal intensity (dots) detected for
increasing input pulse energy (sample of set 1 with $d=6$ u.c.); the
line is a quadratic best-fit.} \label{fig_parabola}
\end{figure}

\begin{figure}[t]
\includegraphics[scale=0.75]{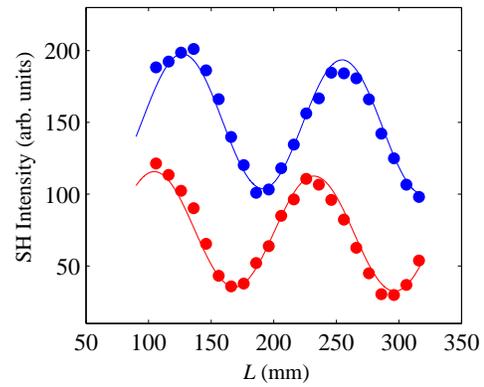}
\caption{(color online). Two examples of SH interference fringes
observed in our HSHG experiments ($sp$ polarizations) for samples of
set 1 having LAO thickness $d=3$ u.c.\ (lower curve, red dots) and
$d=4$ u.c.\ (upper curve, blue dots).} \label{fig_fringes}
\end{figure}

The schematic of our SHG experiment is shown in Fig.\
\ref{fig_experiment}. A Nd:YAG mode-locked laser delivered 20 ps
long pulses at a repetition rate of 10 Hz, which were focused on the
sample with an energy of $\simeq$2 mJ in a spot-area of $\simeq$1
mm$^2$. The input photon energy of 1.17 eV (1064 nm) is well below
the gap energy of both LAO (5.6 eV) and STO (3.3 eV), so that all
possible photoinduced effects are minimized. The SHG intensity
signal from all samples was found to be stable in time and to vary
quadratically with the input laser energy (Fig.\
\ref{fig_parabola}), confirming that the interface properties were
not noticeably altered by the irradiation.
The laser irradiation was also found not to induce any significant
photoconductivity. For our measurements, the SHG beam generated in
reflection from the upper surface of the samples was selected.
Because the LAO film thickness is very small as compared to the
optical wavelength, this SHG signal may include contributions of the
LAO upper surface and of the LAO/STO interface, without significant
propagation-induced phase-shifts between them. For a given optical
geometry, the SHG signal of the entire interfacial region is
determined by its integrated effective nonlinear susceptibility
\begin{equation}
\chi^{(2)}_{\text{eff}}=\int e^{\text{out}}_i L^{\text{out}}_{ii}(z)
\chi^{(2)}_{ijh}(z) L^{\text{in}}_{jj}(z) L^{\text{in}}_{hh}(z)
e^{\text{in}}_j e^{\text{in}}_h dz,
\end{equation}
where $\chi^{(2)}_{ijh}(z)$ are the local second-order nonlinear
susceptibility tensor elements, $L^{\text{in/out}}_{ii}(z)$ the
input/output Fresnel field factors accounting for the optical
propagation, $e^{\text{in/out}}_i$ are the unit vectors of the
input/output polarization directions (sum over repeated indices is
understood), and $z$ a coordinate along the interface
normal.\cite{shen94} Standard SHG measurements give a signal that is
proportional to the squared-modulus $|\chi^{(2)}_{\text{eff}}|^2$.
This quantity, or its square-root $|\chi^{(2)}_{\text{eff}}|$,
provides an estimate of the ``degree of polarity'' of the interface
electrons. More information on the electronic rearrangements, for
example the \emph{direction of the polar asymmetry} (which
determines the sign of the $\chi^{(2)}_{\text{eff}}$), is derived
from the full $\chi^{(2)}_{\text{eff}}$, which in general is a
complex quantity. To measure the $\chi^{(2)}_{\text{eff}}$, we
adopted a homodyne SHG (HSHG) detection geometry \cite{dadap99} as
described in Fig.\ \ref{fig_experiment}. Examples of the resulting
interference fringes are shown in Fig.\ \ref{fig_fringes}. With
suitable fitting,\cite{dadap99} such patterns allowed us to obtain
for each sample both modulus and phase of the complex
$\chi^{(2)}_{\text{eff}}$ (up to a constant phase, which is the same
for all measurements sharing the same experimental geometry). These
measurements were performed for the four input/output polarization
combinations $ss, ps, sp, pp$ (see Fig.\ \ref{fig_experiment}), each
corresponding to a different $\chi^{(2)}_{\text{eff}}$, which in the
following will be respectively denoted as $\chi^{(2)}_{ss},
\chi^{(2)}_{ps}, \chi^{(2)}_{sp}, \chi^{(2)}_{pp}$. All SHG
measurements with $s$ output yielded negligible signals, i.e. we
find $\chi^{(2)}_{ss}\simeq0$ and $\chi^{(2)}_{ps}\simeq0$, as
expected for the C$_{4v}$ symmetry of our interfaces. We note that
both the vanishing of the $s$-polarized SHG and the reproducibility
of our HSHG phase measurements confirm that our signal is coming
from the interface and not resulting from hyper-Rayleigh scattering
from the STO substrate.

\section{Results and discussion}
\begin{figure*}[t]
\begin{center}
\includegraphics[angle=0,scale=0.85]{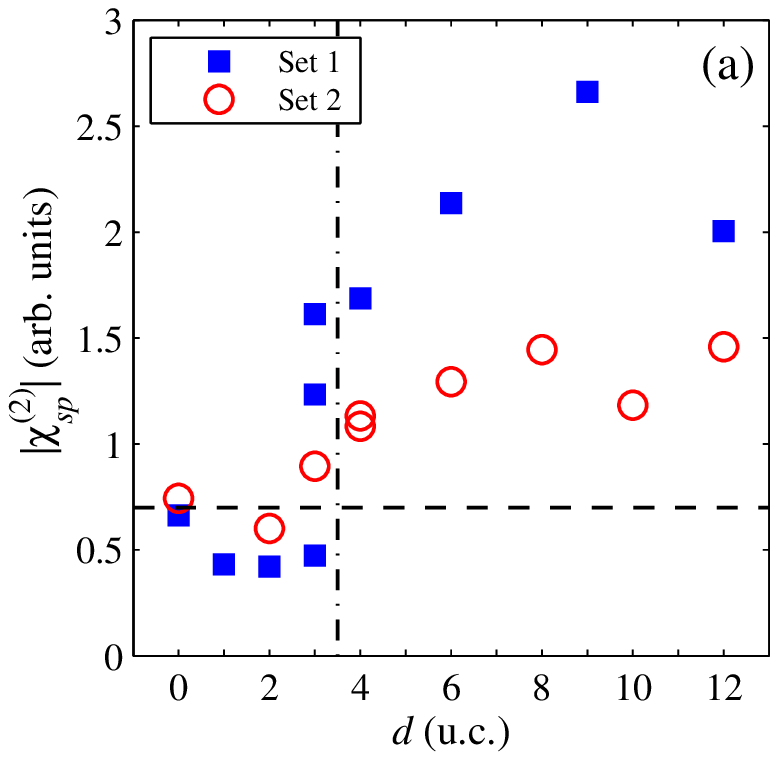}
\hspace{5mm}
\includegraphics[angle=0,scale=0.85]{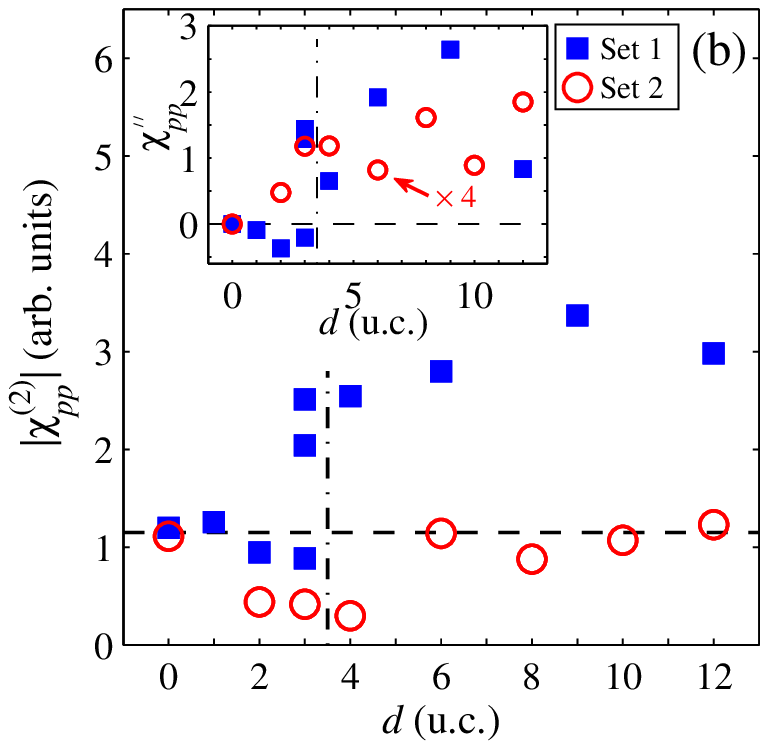}
\par\vspace{3mm}
\includegraphics[angle=0,scale=0.85]{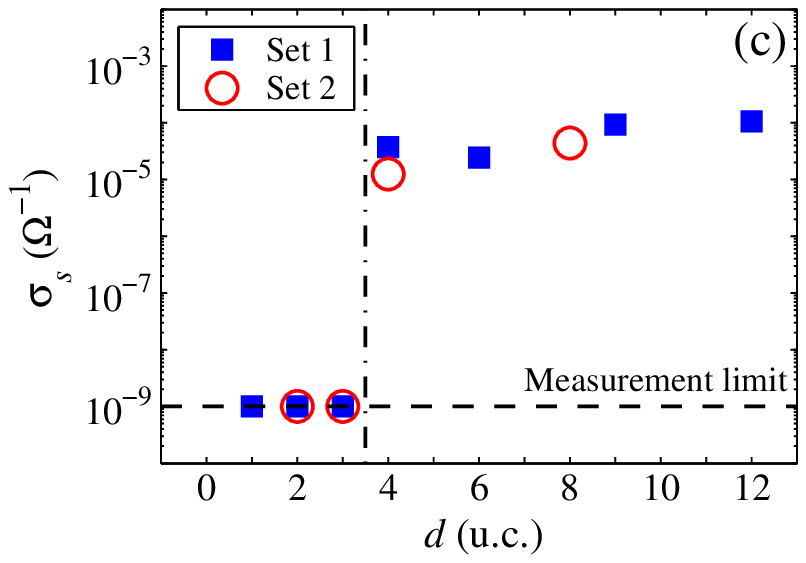}
\hspace{5mm}
\includegraphics[angle=0,scale=0.85]{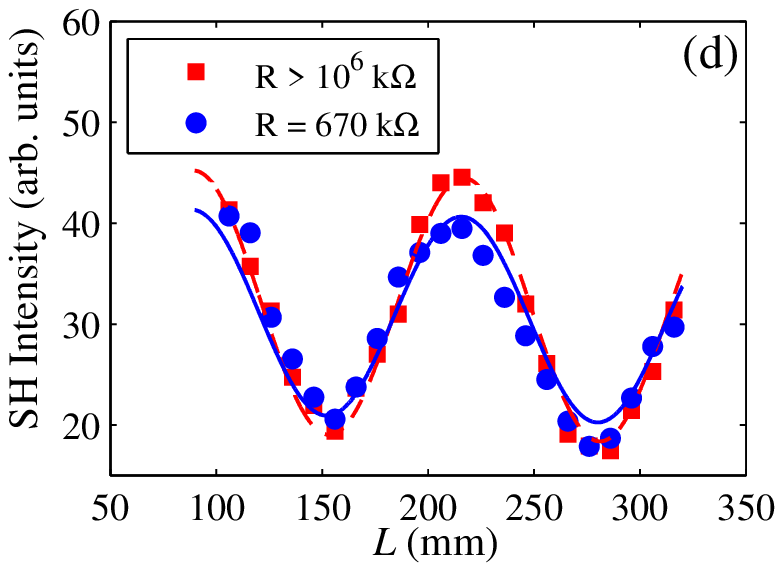}
\end{center}
\caption{(color online). Panels (a)-(b): Amplitude of the SHG
effective nonlinear susceptibilities $\chi^{(2)}_{sp}$ (a) and
$\chi^{(2)}_{pp}$ (b) measured as a function of LAO thickness $d$
for samples of set 1 (blue squares) and set 2 (red circles).
Different data points with the same $d$ refer to different samples.
The dashed horizontal line gives the average SHG amplitude of the
STO substrates ($d=0$). The dot-dashed vertical line corresponds to
the measured threshold thickness (i.e., between 3 and 4 u.c.) for
the onset of conductivity. Inset of panel (b): imaginary component
of $\chi^{(2)}_{pp}$ (data of set 2 are rescaled by a factor 4 for
clarity). Panel (c): Sheet conductivity measured for most of our
samples; note that all four samples with $d=3$ u.c.\ were found to
be insulating, in the absence of external fields. Panel (d): HSHG
$sp$ signal for a $d=3$ u.c.\ sample of set 2 in its insulating (red
squares) and conductive (blue circles) state, respectively obtained
by applying $-100$ V and $+100$ V to a back-gate.\cite{thiel06} All
data were taken at room temperature.} \label{fig_ampl}
\end{figure*}

Figures \ref{fig_ampl}a-b show the amplitude (modulus) of the $sp$
and $pp$ effective nonlinear susceptibilities measured as a function
of the LAO film thickness $d$. We note that STO substrates ($d=0$)
already generate a significant signal. In contrast, we measured a
negligible SHG from the (001) surface of a LAO single crystal. In
LAO/STO heterostructures, the SH amplitude $|\chi^{(2)}_{sp}|$ is
seen to be approximately constant for $0\leq d \leq 2$ u.c.\, with
only a slight decrease observed for samples having 1 or 2 monolayers
of LAO with respect to bare STO substrates (Fig.\ \ref{fig_ampl}a).
When the LAO thickness reaches $d=3$ u.c., however, an abrupt and
substantial increase of the SHG intensity takes place. The SHG
signal obtained for different 3 u.c.\ samples also exhibits a strong
scatter, that is not seen for other thicknesses. Both observations
indicate that $d=3$ u.c.\ is the threshold value for a discontinuous
structural transition. For larger $d$, the SHG amplitude increases
further to saturate or to decrease again for $d\gtrsim10$ u.c.\
(with some scatter from sample to sample). This step-like behavior
is clearly evident for the samples of set 1. For samples of set 2
the signal is smaller and the transition seems more gradual. This
fact implies that more electrons are involved in the interfacial
process for samples of set 1 as compared to those of set 2, in
accordance with the difference of the measured carrier densities. In
the $pp$ geometry (Fig.\ \ref{fig_ampl}b), samples of set 1 behave
quite similarly to the $sp$ case, while samples of set 2 show a more
complex behavior, with a decrease of SH intensity at $d=2$ u.c.\ and
an increase at $d=6$ u.c. In this case, however, a step-like
behavior of the imaginary component of $\chi^{(2)}_{pp}$ is still
seen (inset of Fig.\ \ref{fig_ampl}b).

The measured behavior of the SHG is reminiscent of the abrupt
conductance change that is found in the samples as a function of $d$
(Fig.\ \ref{fig_ampl}c). However, the conduction step occurs for
$d\geq4$ u.c., while the SHG step has been found to take place for
$d\geq3$ u.c. This implies that the SHG signal is not detecting
directly the mobile electrons, but it is instead \emph{revealing a
related phenomenon that acts as a precursor for the onset of
conductivity}. This important conclusion is further confirmed by the
observation that the HSHG signal is not sensitive to the switch in
conductivity that can be induced in a $d=3$ u.c.\ sample by an
applied back-gate voltage (Fig.\ \ref{fig_ampl}d). On the other
hand, SHG is not expected to be specifically sensitive to the
conduction itself. An increase of the SHG amplitude may reflect, in
general, either an increase of polarizing electric fields
experienced by the interfacial electrons (possibly also reflecting
lattice distortions) or a transfer of electrons from less
polarizable and/or less polar orbitals to more polarizable and/or
more polar ones.

\begin{figure}[t]
\begin{center}
\includegraphics[scale=0.8, bb= 138 319 280 548,
clip=true]{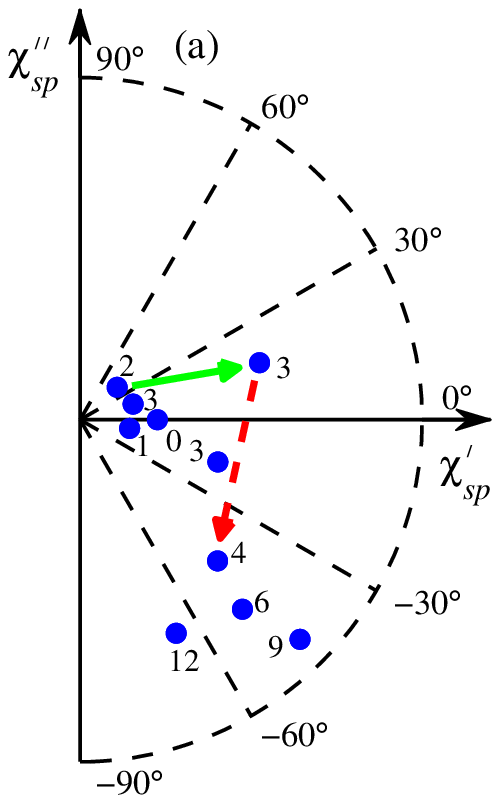}
\includegraphics[scale=0.8, bb= 144 319 288 548,
clip=true]{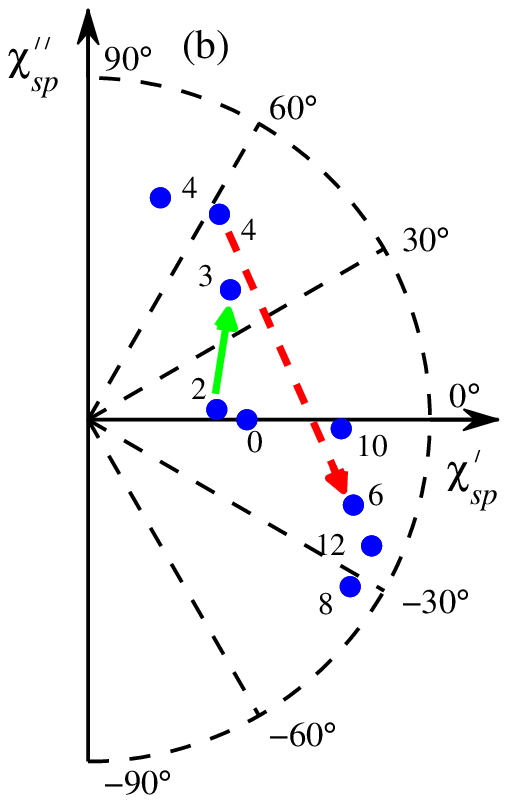}
\par\vspace{5 mm}
\includegraphics[scale=0.8, bb= 137 319 279 549,
clip=true]{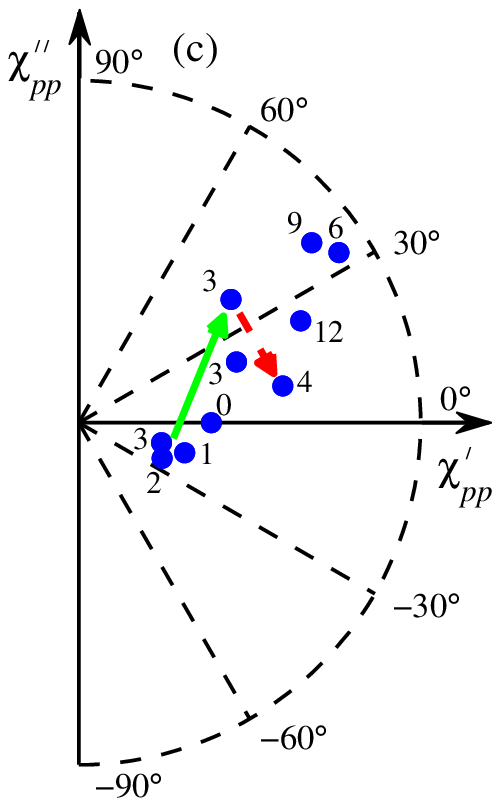}
\includegraphics[scale=0.75, bb= 131 299 284 548,
clip=true]{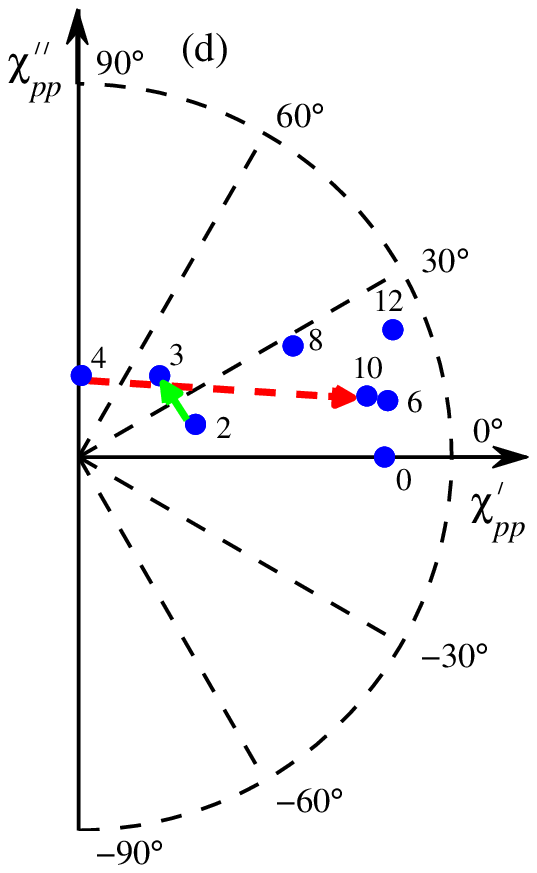}
\end{center}
\caption{(color online). Complex effective SHG nonlinear
susceptibility $\chi^{(2)}_{sp}=\chi'_{sp}+i\chi''_{sp}$ (panels a
and b) and $\chi^{(2)}_{pp}=\chi'_{pp}+i\chi''_{pp}$ (panels c and
d) of the LAO/STO heterostructure for samples of set 1 (a and c) and
set 2 (b and d) having different LAO thicknesses $d$. The polar
angle of each point corresponds to the argument (or phase) of the
complex susceptibility, as measured by HSHG (defined to zero for
$d=0$). The numbers typed next to the data points give the
thicknesses of the LAO films in u.c. The two arrows in each panel
(solid-line green and dashed-line red) indicate the two abrupt
electronic transitions discussed in the text.} \label{fig_polar}
\end{figure}

The full complex nonlinear susceptibility $\chi^{(2)}_{\text{eff}}$
provides further useful information on the electronic behavior of
the interfaces. It is convenient to present these data in a complex
plane: in this representation, the extent of the electronic
rearrangements that result from each addition of a monolayer of LAO
is directly related to the distance between consecutive data points
in the plane. As Fig.\ \ref{fig_polar} shows, also the data of
complex susceptibility exhibit large variations at $d=3$ u.c.\
(green solid-line arrows), but small ones for thinner LAO layers.
For larger $d$, however, another abrupt and large variation of the
HSHG signal is found (red dashed-line arrows in Fig.\
\ref{fig_polar}). Because this is mainly a phase variation, it is
not well visible in the amplitude plots discussed above. Such a
phase shift in the nonlinear susceptibility can only result from an
electronic transfer, as optical phase retardations are determined by
the optical resonances of the electronic polarizability. Samples in
set 1 (Fig.\ \ref{fig_polar}a,c) exhibit this second transition
between $d=3$ u.c.\ and $d=4$ u.c.\ (see also Fig.\
\ref{fig_fringes}), i.e., in coincidence with the onset of
conduction. The samples of set 2 (Fig.\ \ref{fig_polar}b,d) also
show this second transition. Yet, they behave a bit differently: the
change of $\chi^{(2)}_{\text{eff}}$ at $d=4$ u.c.\ is small, and the
second transition is seen only when passing from 4 u.c.\ to 6 u.c.
Also this difference in the behavior of the two sample sets is
likely related with the larger overall density of electrons involved
in the interfacial process for set 1, leading, at this second
transition, to a ``faster'' variation with $d$.

\section{Possible interpretations}
Two main alternative scenarios are seen as candidates of the
electronic effects underlying the two transitions observed with
HSHG.

Scenario 1 is based on the assumption that the SHG signal is
dominated by the electronic states residing in the STO because its
electric and optical polarizabilities far exceed those of LAO. This
assumption is also suggested by the observation of a negligible SHG
from the LAO single crystal. In this scenario, at a LAO thickness of
3 u.c. the ``polar catastrophe'' begins and electrons start to be
injected from LAO into STO interface states. Because interfaces with
$d=3$ u.c.\ are insulating, these electrons must be trapped in
localized surface states. Although not mobile, these electrons
provide the main contribution to the first SHG transition. The
possible presence of localized electrons at the LAO/STO interface
has been addressed in several studies, in connection with
strong-correlation effects, lattice deformations, self-trapped
polarons, etc. (see, e.g., Refs.\
\onlinecite{pentcheva06,vonk07,rubano07,zhong08,pentcheva08}), and
may be related to the magnetic effects seen in suitable conditions
at low temperatures.\cite{brinkman07} A particularly intriguing
possibility is that the trapping mechanism is a form of
disorder-driven Anderson localization taking place in the
quasi-two-dimensional electron gas.\cite{popovic08} Recent evidence
of extreme sensitivity of carrier concentrations to relatively small
interface delta-doping favors this hypothesis.\cite{fix09} Disorder
at $d=3$ u.c.\ may be strongly enhanced by the intrinsic electronic
bistability of the system,\cite{cen08,cen09} which might be also
reflected in the large sample-to-sample variability observed in our
SHG signal. It is also possible that such bistability gives rise to
a phase-separation at the interface, with non-percolating conducting
regions surrounded by insulating ones. On further increasing the LAO
film thickness, more electrons are injected and give rise to
conduction. The onset of conduction may be related with a reduced
disorder (as bistability is not observed for $d\geq4$ u.c.) or it
may arise because electrons start to occupy higher-energy
interfacial orbitals, e.g., different Ti $3d$-$t_{2g}$
subbands,\cite{popovic08,salluzzo09,copie09} including extended
states. Within this scenario, we would ascribe, in particular, the
second SHG transition seen in the polar plots of
Fig.~\ref{fig_polar} to the filling of higher-energy interfacial
subbands. Our SHG findings would then be consistent with the
localized-electrons explanation of the missing charge
problem.\cite{popovic08}

Scenario 2 assumes that the contribution to SHG of the LAO orbitals
is not negligible and that at $d=3$ u.c.\ the polarity of the LAO
layer suddenly increases. The electronic polarity of thinner LAO
films may be depressed, e.g., by interfacial roughness, cationic
mixing,\cite{nakagawa06,willmott07} or lattice
distortions.\cite{vonk07,pentcheva09,jia09} These effects could be
particularly large in the two outer monolayers of the LAO film that
are adjacent to the STO and to the air, possibly explaining the 3
u.c.\ threshold. \emph{Ab initio} calculations of the ionic
relaxations taking place in the LAO film support in part this
concept, as the ionic relaxation is indeed predicted to be weak in
the outermost atomic planes of LAO.\cite{pentcheva09} In this
scenario, only the second transition seen by SHG is ascribed to the
LAO$\rightarrow$STO electronic injection. We believe this second
scenario to be less likely then the first one, but it cannot be
excluded at present.

\section{Conclusions}
In summary, the interfacial reconstructions taking place in LAO/STO
heterostructures as a function of the LAO thickness have been
investigated by using optical second harmonic generation. Two
distinct electronic transitions were found, which result from the
reorganization of the electrons at the LAO/STO interface induced by
the polar discontinuity. In the most plausible interpretation
scenario, the second-harmonic signal provides evidence that, at the
critical LAO thickness of 3 unit cells, electrons are already
injected in the interface but become localized. This might be linked
to the recently demonstrated electronic bistability of the interface
at this critical thickness, which can give rise to disorder-driven
Anderson localization or, possibly, a complete phase separation
between conducting and insulating areas. For higher LAO thickness
the number of injected electrons increases and their distribution
becomes more uniform, thus giving rise to the observed conduction.
In addition, evidence for the existence of distinct interfacial
electronic subbands is provided by the second transition in the
optical signal seen for increasing LAO thickness.

\section*{Acknowledgments}
We thank Daniele Marr\'e, Ilaria Pallecchi, and Marta Codda for Hall
measurements, Davide Maccariello  for help in the sample growth, and
Marco Siano and Romolo Savo for help in some SHG measurements. This
work was supported by the EU (Nanoxide) and by the DFG (SFB484).


\end{document}